\documentclass[%
reprint,
superscriptaddress,
 amsmath,amssymb,
 aps,
prd,
floatfix,
]{revtex4-2}
\usepackage{graphicx}
\usepackage{dcolumn}
\usepackage{bm}
\usepackage{xcolor}
\usepackage{xspace}
\usepackage{subfigure}
\usepackage{topcapt}
\usepackage{float}
\usepackage{slashed}

\newcommand{\GEANTfour}{{\textsc{Geant4}}\xspace}

\newcommand{\Npe}{\ensuremath{n_\text{pe}}\xspace}

\newcommand{\mCP}{\ensuremath{\chi}\xspace}

\begin{document}


\title{Sensitivity to millicharged particles in future proton-proton collisions at the LHC with the milliQan detector}

\author{A.~Ball}\affiliation{CERN, Geneva, Switzerland}
\author{J.~Brooke}\affiliation{University of Bristol, Bristol, United Kingdom}
\author{C.~Campagnari}\affiliation{University of California, Santa Barbara,  California 93106, USA}
\author{M.~Carrigan}\affiliation{The Ohio State University, Columbus, Ohio 43218, USA}
\author{M.~Citron}\affiliation{University of California, Santa Barbara,  California 93106, USA}
\author{A.~De~Roeck}\affiliation{CERN, Geneva, Switzerland}
\author{M.~Ezeldine}\affiliation{Lebanese University, Hadeth-Beirut, Lebanon}
\author{B.~Francis}\affiliation{The Ohio State University, Columbus, Ohio 43218, USA}
\author{M.~Gastal}\affiliation{CERN, Geneva, Switzerland}
\author{M.~Ghimire}\affiliation{New York University, New York, New York 10012, USA}
\author{J.~Goldstein}\affiliation{University of Bristol, Bristol, United Kingdom}
\author{F.~Golf}\affiliation{University of Nebraska, Lincoln, Nebraska 68588, USA}
\author{A.~Haas}\affiliation{New York University, New York, New York 10012, USA}
\author{R.~Heller}\thanks{Now at Fermi National Accelerator Laboratory, Batavia, Illinois 60510, USA.}\affiliation{University of California, Santa Barbara,  California 93106, USA}
\author{C.S.~Hill}\affiliation{The Ohio State University, Columbus, Ohio 43218, USA}
\author{L.~Lavezzo}\affiliation{The Ohio State University, Columbus, Ohio 43218, USA}
\author{R.~Loos}\affiliation{CERN, Geneva, Switzerland}
\author{S.~Lowette}\affiliation{Vrije Universiteit Brussel, Brussels 1050, Belgium}
\author{B.~Manley}\affiliation{The Ohio State University, Columbus, Ohio 43218, USA}
\author{B.~Marsh}\affiliation{University of California, Santa Barbara,  California 93106, USA}
\author{D.W.~Miller}\affiliation{University of Chicago, Chicago, Illinois 60637, USA}
\author{B.~Odegard}\affiliation{University of California, Santa Barbara,  California 93106, USA}
\author{R.~Schmitz}\affiliation{University of California, Santa Barbara,  California 93106, USA}
\author{F.~Setti}\affiliation{University of California, Santa Barbara,  California 93106, USA}
\author{H.~Shakeshaft}\affiliation{CERN, Geneva CH-1211, Switzerland}
\author{D.~Stuart}\affiliation{University of California, Santa Barbara,  California 93106, USA}
\author{M.~Swiatlowski}\thanks{Now at TRIUMF, Vancouver BC V6T 2A3, Canada.}\affiliation{University of Chicago, Chicago, Illinois 60637, USA}
\author{J.~Yoo}\thanks{Now at Korea University, Seoul 02841, South Korea.}\affiliation{University of California, Santa Barbara,  California 93106, USA}
\author{H.~Zaraket}\affiliation{Lebanese University, Hadeth-Beirut, Lebanon}

\date{\today}

\begin{abstract}
\noindent We report on the expected sensitivity of dedicated scintillator-based detectors at the LHC for elementary particles with charges much smaller than the electron charge.
The dataset provided by a prototype scintillator-based detector is used to characterise the performance of the detector and provide an accurate background projection. Detector designs, including a novel slab detector configuration, are considered for the data taking period of the LHC to start in 2022 (Run 3) and for the high luminosity LHC. With the Run 3 dataset, the existence of new particles with masses between 10~MeV and 45~GeV could be excluded at 95\% confidence level for charges between 0.003e and 0.3e, depending on their mass. With the high luminosity LHC dataset, the expected limits would reach
between 10~MeV and 80~GeV 
for charges between 0.0018e and 0.3e, depending on their mass.
\end{abstract}

\maketitle

\section{\label{sec:intro}Introduction}

The nature of the dark matter (DM) remains one
of the most compelling unanswered questions
in particle physics.  For a long time the focus
of theoretical and experimental investigations
has been on models where all DM is composed of
a single particle~\cite{Bertone_2005}.
More recently, theories
where the DM is comprised of a set of particles
with their own ``dark" interactions have gained
prominence~\cite{Battaglieri:2017aum}. 

One can consider a dark sector containing a massless abelian gauge field, $A'$, that couples to a new dark fermion, $\chi$, with order one coupling, $e'$. A kinetic mixing, $\kappa$, can be introduced between the $A'$ and SM hypercharge $B$. Under a convenient basis, $A'$ is decoupled from the SM sector and the Lagrangian can be written as

\begin{equation*}
   \mathcal{L}_{\rm dark} \subset -\frac{1}{4}A'_{\mu\nu}A^{\prime\mu\nu} + i\bar{\chi}\left( \slashed{\partial} + ie' \slashed{A}' - i \kappa e' \slashed{B} + im_{\chi} \right)\chi .\
\label{eqn:mCP_lag} 
\end{equation*}

In this case, the $\chi$ acts as a field with hypercharge $\kappa e'$~\cite{Holdom:1985ag, Izaguirre:2015eya}. The new fermion is generically called a millicharged particle since a natural value for $\kappa$, and therefore the $\chi$ effective electric charge, of $\sim10^{-3}$ arises from one-loop effects~\cite{Davidson:1993sj}. 

While direct searches robustly probe the parameter space of millicharged particles, constraints from indirect observations can be evaded by adding degrees of freedom, which can readily occur in minimally extended dark sector models~\cite{Izaguirre:2015eya}.  In particular, the parameter space 1 $ < m_\mCP < 100$~GeV, an ideal mass range for production at the LHC, is largely unexplored by direct searches~\cite{MilliQ,essig2013dark,Chatrchyan_2013,Chatrchyan_2013_2,Acciarri_2020, Davidson:2000hf, Badertscher:2006fm,PhysRevLett.122.071801,Marocco_2021,Plestid_2020}. 

In response to this, some of us discussed the possibility to build the ``milliQan" experiment in the PX56 drainage and observation gallery located at LHC P5 pointing to the CMS interaction point (IP)~\cite{Ball:2016zrp}. As detailed in Ref.~\cite{ball2020search}, we have installed and operated a small fraction of such a detector (``milliQan demonstrator'').  With data from the demonstrator we have already excluded some of the previously unconstrained parameter space for millicharged particles.  More importantly,
this prototype detector provided crucial insights into the dominant sources  of backgrounds, the efficiency of detection for millicharge signals, and the design of other milliQan-like experiments proposed at accelerator facilities around the world~\cite{fermini,formosa,submet, moedal}. 

Having secured the necessary funding, we are preparing to install two complementary detectors at the P5 experimental site
for the data taking period of the LHC starting in 2022 (Run 3): a ``bar" detector upgrade of the milliQan demonstrator and a novel ``slab" detector design. In this paper, we provide prospects for these detectors, as well as for an extension of the 
design for the high-luminosity LHC (HL-LHC), demonstrating achievable sensitivities for millicharged particles of 
masses in the range 10~MeV to 80~GeV with $Q \sim$ 0.0018 to 0.3e.

\section{\label{sec:detector} Detector design}

As detailed in Ref.~\cite{ball2020search}, the milliQan experimental cavern is located in an underground tunnel at a distance of 33~m from the CMS IP, with 17~m of rock between the IP and the detector that provides shielding from most particles produced in LHC collisions. In order to be sensitive to particles with charges as low as $0.001$e a large active area of scintillator is required. For Run 3, two detector designs are planned for deployment: a bar detector and a slab detector. In the CMS coordinate system~\cite{CMSTDR}, the bar detector will be positioned at an azimuthal angle ($\phi$) of $43^\circ$ and pseudorapidity ($\eta$) of $0.1$. The slab detector will be placed around 5~m behind the bar detector at $\phi = 38^\circ$, a distance of 37~m from the IP.

\begin{figure}[!htp]
    \includegraphics[width=.8\linewidth]{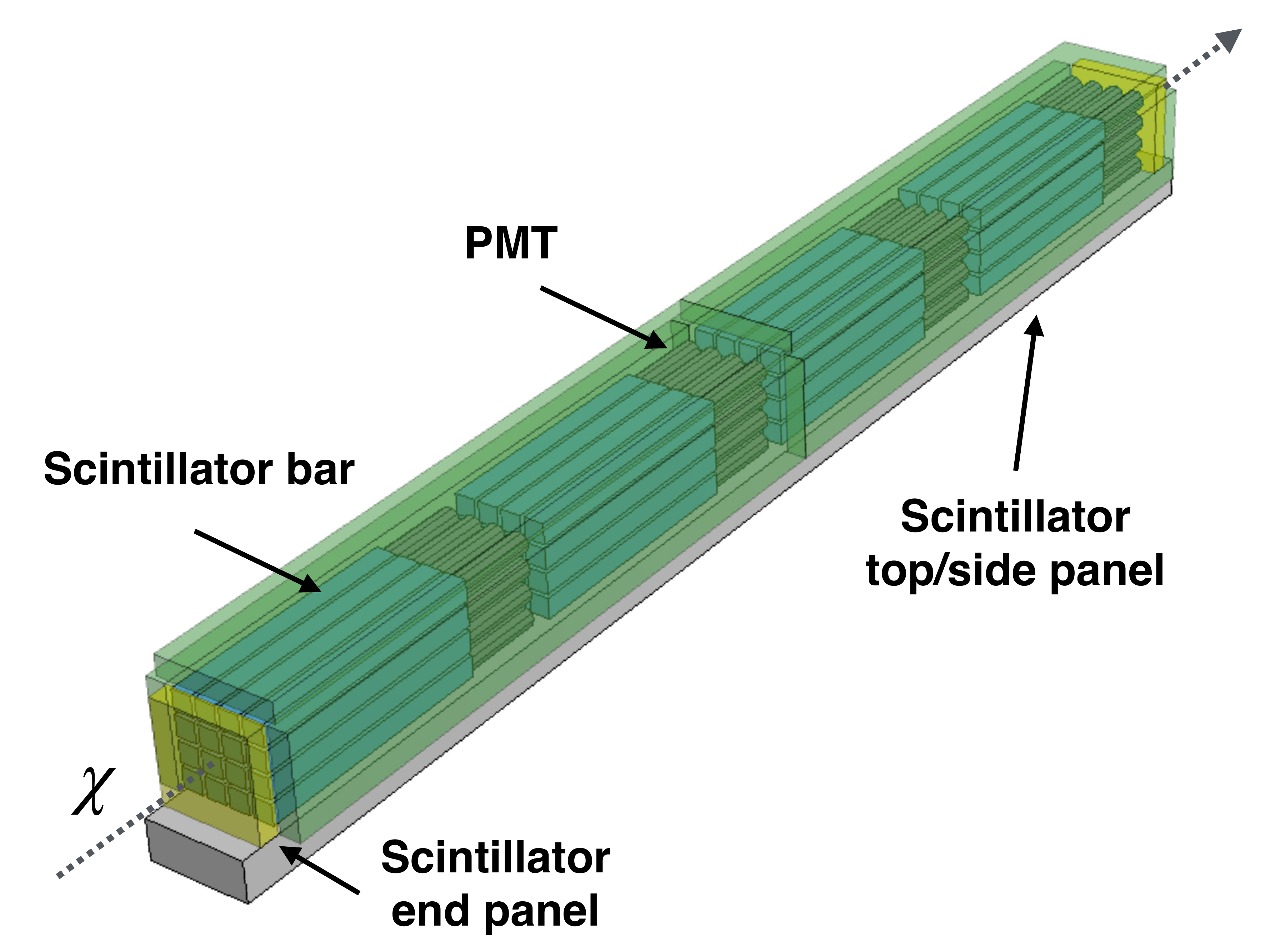}
    \caption{\protect A diagram of the milliQan Run 3 bar detector components. The scintillator bars are shown in blue connected to PMTs in black. The side and top panels are shown surrounding the bars in transparent green while the end panels are shown in transparent yellow. The PMTs are  not  shown  for  the  side and top panels. All components are installed on an aluminum tube. The path of a millicharged particle from the IP is shown in gray.}
    \label{fig:barDetFig}
\end{figure}

The Run 3 bar detector is comprised of a $0.2~\mathrm{m}~\times0.2~\mathrm{m}~\times3~\mathrm{m}$ plastic scintillator array. 
The array will be oriented such that the long axis points at the nominal CMS IP. The array will contain four longitudinal ``layers", each containing sixteen $5~\mathrm{cm}\times5~\mathrm{cm}\times60~\mathrm{cm}$ scintillator ``bars" optically coupled to high-gain photomultiplier tubes (PMTs) in a $4\times4$ array. Surrounding the array is an active muon veto shield composed of six 5 cm thick scintillator panels that cover the top and sides of the array. Each panel will have two PMTs at opposing ends to increase light collection efficiency and to provide some positional information (using relative pulse sizes and $\sim$ns timing resolution). An additional small scintillator panel at each end of the bars will complete the hermeticity of the shield. These end panels will also be used to discriminate higher charge signals from the deposits of muons originating at the CMS IP using the pulse size, as in Ref.~\cite{ball2020search}. A diagram of the bar detector may be seen in Fig.~\ref{fig:barDetFig}. 
 
The bar detector design closely follows the design of the milliQan demonstrator, with several important upgrades. These are an increase in surface area from 150~$\textrm{cm}^2$ to 400~$\textrm{cm}^2$, the addition of a fourth layer for improved background rejection, an increase in the scintillator veto panel thickness from 0.5~$\textrm{cm}$ to 5~$\textrm{cm}$, the inclusion of an amplifier attached to the readout of each PMT to allow single photoelectron pulses to be reconstructed with near 100\% efficiency, and an LED flasher system for calibration and monitoring. The LEDs will be used to measure the average area of single photoelectron waveforms for each channel, following the method outlined in Ref.~\cite{Saldanha:pmt}. The response for millicharged particles will be calibrated using the measured area of known energy depositions from a range of radioactive sources as well as cosmic muons.

As will be shown in Section~\ref{sec:proj}, the sensitivity for a \mCP with mass above $\sim1.4$ GeV is limited by the angular acceptance of the detector and not the efficiency of the scintillator bars. This motivates an additional detector that makes use of a large active area of thinner scintillator, the ``slab detector". While the thinner scintillator results in a reduction in sensitivity at the smallest charges, its expanded geometric coverage allows the slab detector to improve the reach for higher \mCP masses.

\begin{figure}[!htp]
    \includegraphics[width=.8\linewidth]{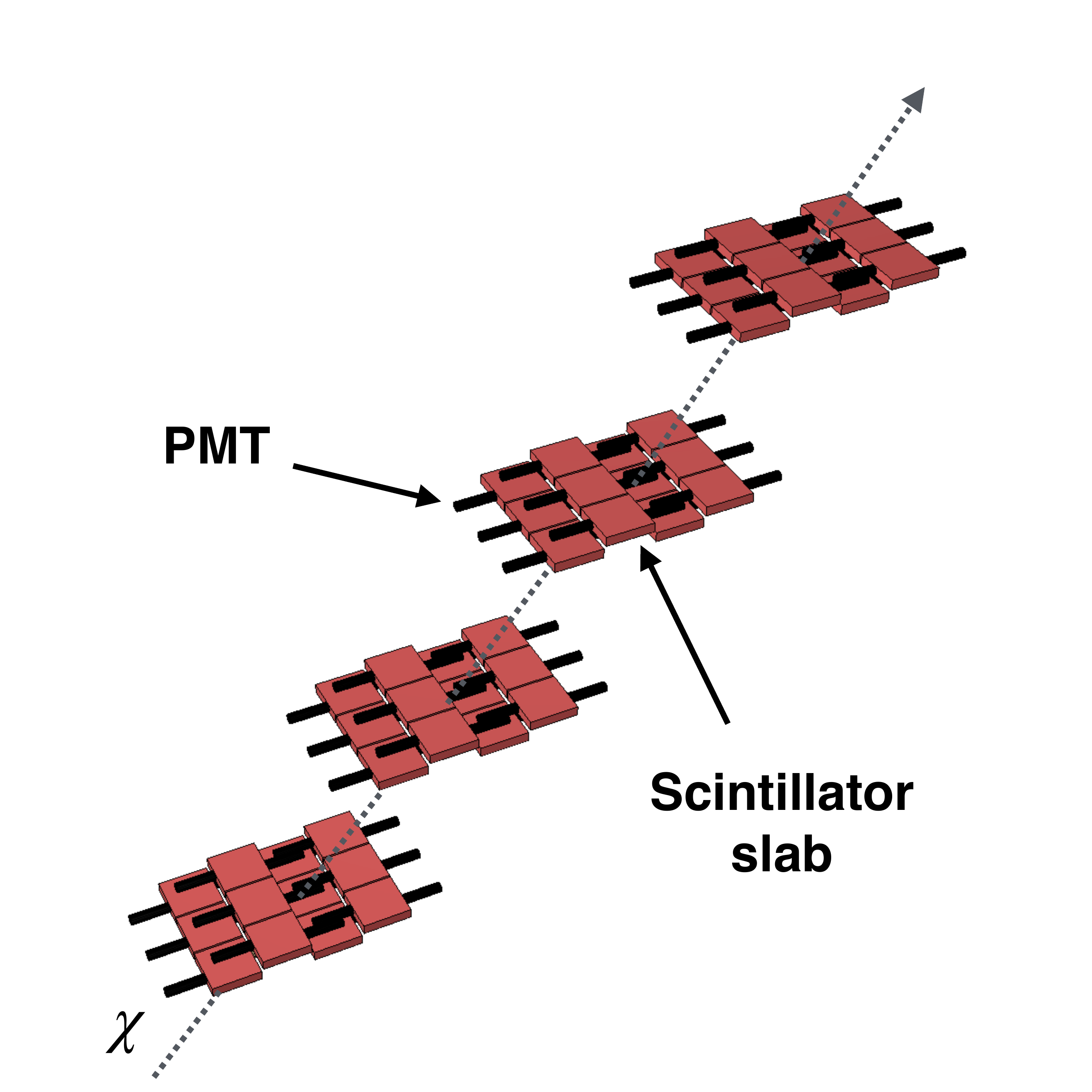}
    \caption{\protect A diagram of the milliQan slab detector components. The scintillator slabs are shown in red connected to PMTs in black. The support structure is not shown. The path of a millicharged particle from the IP is shown in gray.}
    \label{fig:slabDetFig}
\end{figure}

The slab detector will be comprised of $40~\mathrm{cm}\times60~\mathrm{cm}\times5~\mathrm{cm}$ scintillator ``slabs''. These will be arranged in four layers of $3 \times 4$ slabs. There are therefore a total of 48 slabs in the array. The segmentation of the layers in the slab detector
is driven by a compromise between practical considerations, including mechanical constraints and limiting the number of channels, as well as the desire to sharply define pointing paths
to the IP to reduce accidental backgrounds. Each layer of the slabs will be held by a simple shelving unit. A drawing of the slab detector may be seen in Fig.~\ref{fig:slabDetFig}.  Similarly to the Run 3 bar detector, an LED flasher system will be installed, and radioactive sources and muons used to calibrate the response.

For the HL-LHC, should additional funding become available, we consider an upgraded bar detector design.  This detector would be composed of a $1~\mathrm{m}\times1~\mathrm{m}\times3~\mathrm{m}$ plastic scintillator array. 
The arrays are subdivided into nine steps, stacked on top of each other, held in place by a mechanical cage supported by a rotatable mechanical structure. Each step contains six modules in four longitudinal layers, each containing four $5~\mathrm{cm}\times5~\mathrm{cm}\times60~\mathrm{cm}$ scintillator bars, in a $2\times2$ array. There are thus a total of 864 ($9\times6\times4\times4$) bars in the array. The detector is hermetically surrounded by 5~cm thick veto panels on each side and each end.

\section{\label{sec:mc} Event Generation and Simulation}

The basic principles of the Monte Carlo generation and simulation of signals and backgrounds are detailed in Ref.~\cite{ball2020search}.  Briefly, pairs of millicharged particles of spin $\frac{1}{2}$ are generated at $\sqrt{s} = 13$ TeV from modified Standard Model processes such as Drell-Yan, decays of vector mesons, and Dalitz-decays of light mesons.  These particles are transported through the CMS magnetic field and the rock in the cavern to the drainage tunnel where the milliQan detector is installed.  The response of the detector and the readout electronics is modeled with a combination of \GEANTfour~\cite{AGOSTINELLI2003250},
test data from cosmic rays, and bench tests with an LED flasher.  

The understanding of backgrounds arising from cosmic muons that shower in the rock and detector material (``shower'' background) is crucial for the detector design and to estimate the expected sensitivity of the proposed 
detectors. The shower background is estimated
from simulation.  The simulation is validated with data taken with the three-layer demonstrator 
reconfigured in a horizontal position in order to be able to place two additional bars 
at its end to form a (partial) four-layer detector.

A sample of $7.7 \times 10^5$ cosmic triggers were collected with the four-layer demonstrator
in a beam-off period of 1800 hours.  The \GEANTfour based simulated cosmic data set is normalized to the number of data triggers, yielding 
a cosmic flux consistent with the measurements in Ref.~\cite{Bogdanova_2006}.
The probability of multiple cosmic ray muon events is taken into account
in the simulation~\cite{cosmicshowers}.

\begin{figure}[ht]
  \centering
  \includegraphics[width=\linewidth]{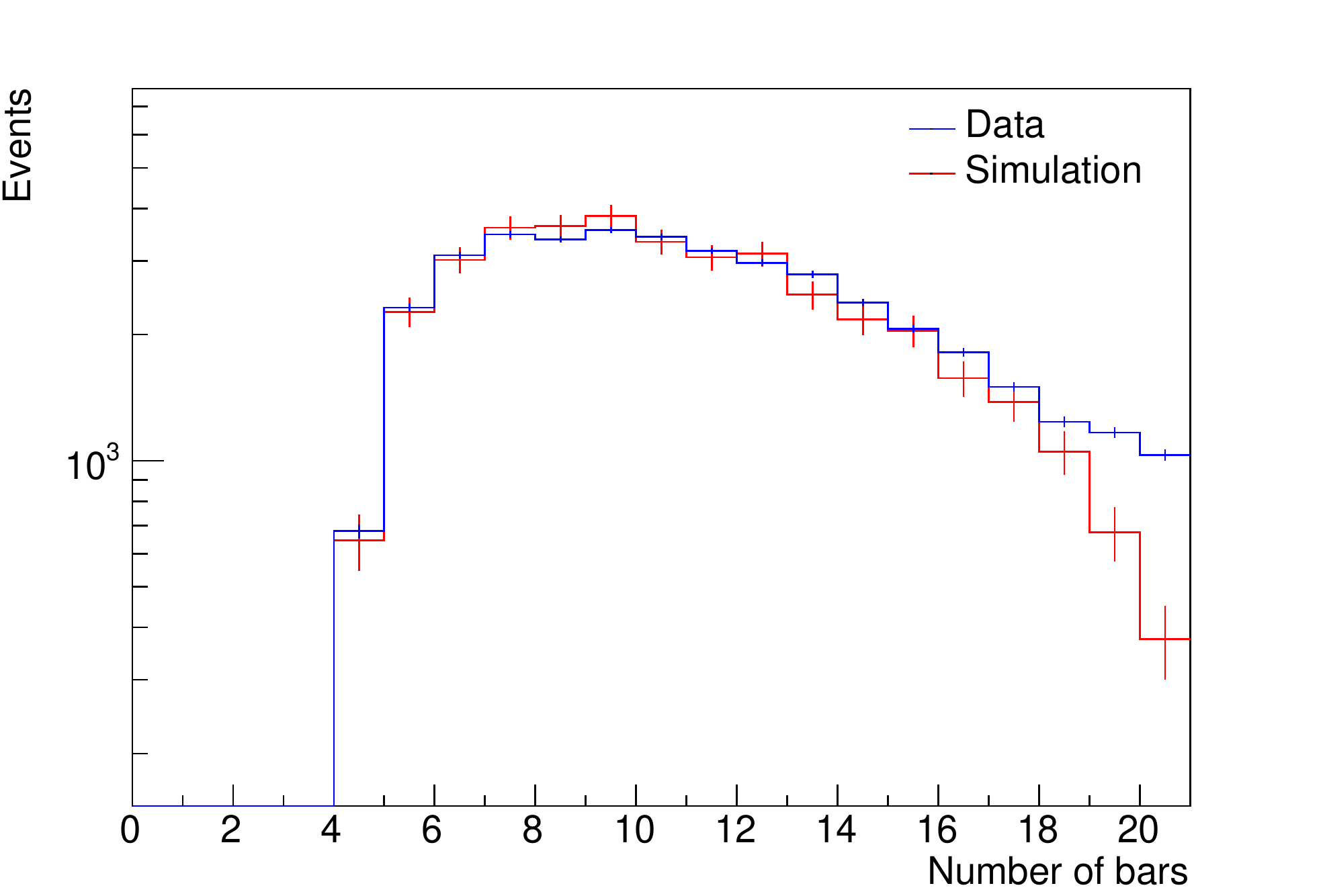}
  \caption{The number of scintillator bars with a detected pulse in cosmic muon events for data (blue) and simulation (red).}
  \label{fig:cosmicSimNbar}
\end{figure}

\begin{figure}[ht]
  \centering
  \includegraphics[width=\linewidth]{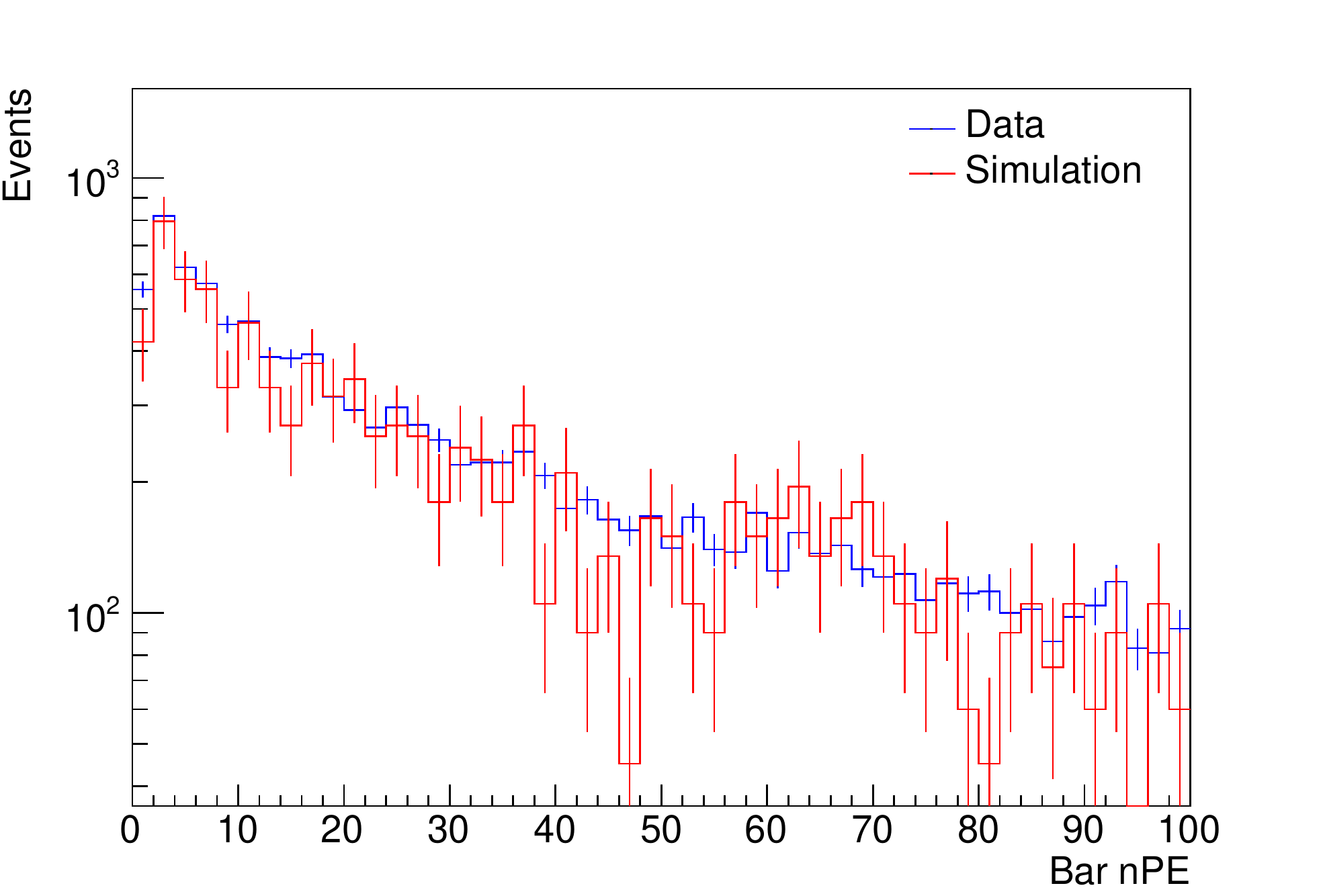}
  \caption{The \Npe for each pulse in cosmic muon events for data (blue) and simulation (red).}
  \label{fig:cosmicSimNPE}
\end{figure}

\begin{figure}[ht]
  \centering
  \includegraphics[width=\linewidth]{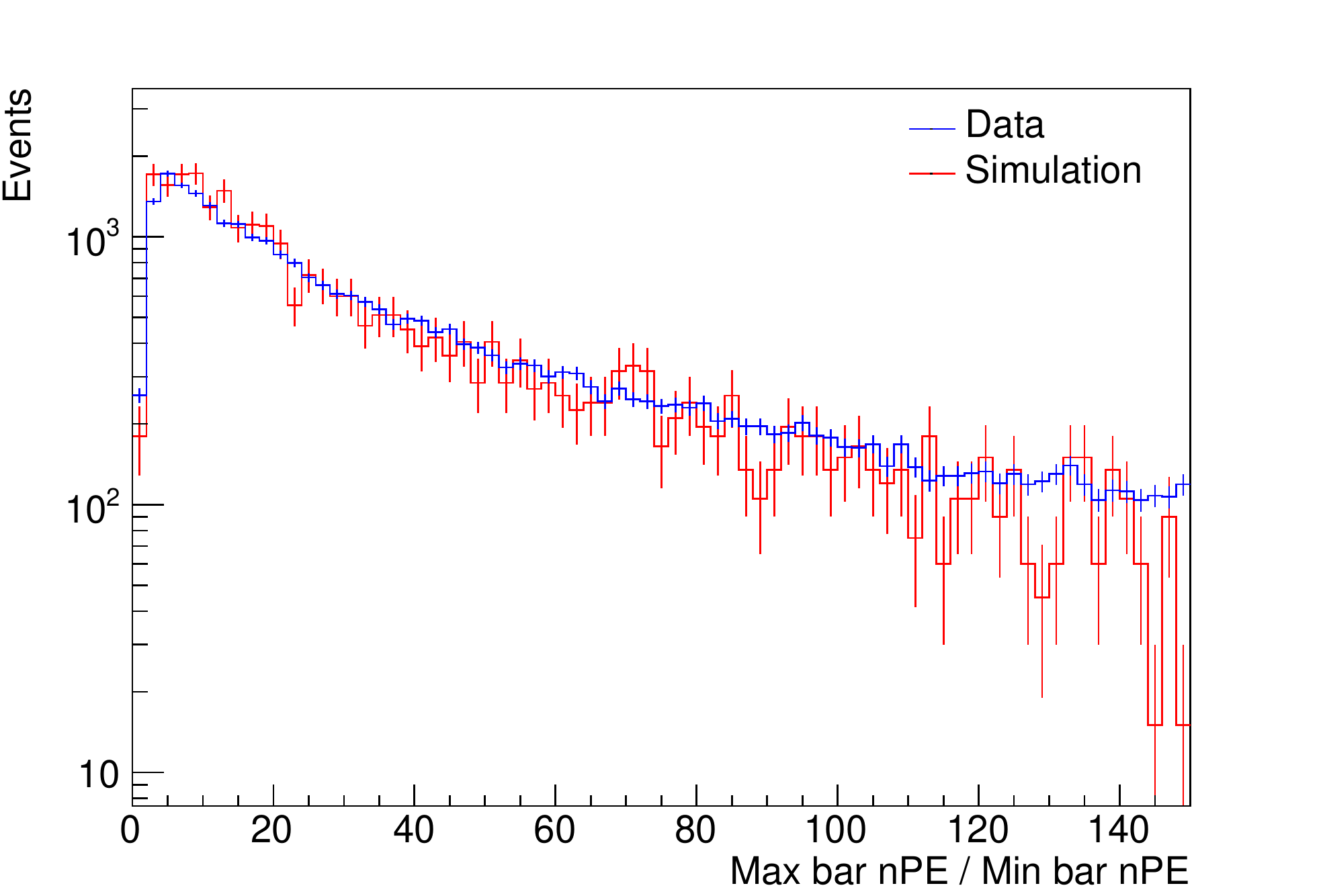}
  \caption{The ratio of maximum \Npe to minimum \Npe in cosmic muon events for  data (blue) and simulation (red).}
  \label{fig:cosmicSimMaxMinNPE}
\end{figure}

\begin{figure}[ht]
  \centering
  \includegraphics[width=\linewidth]{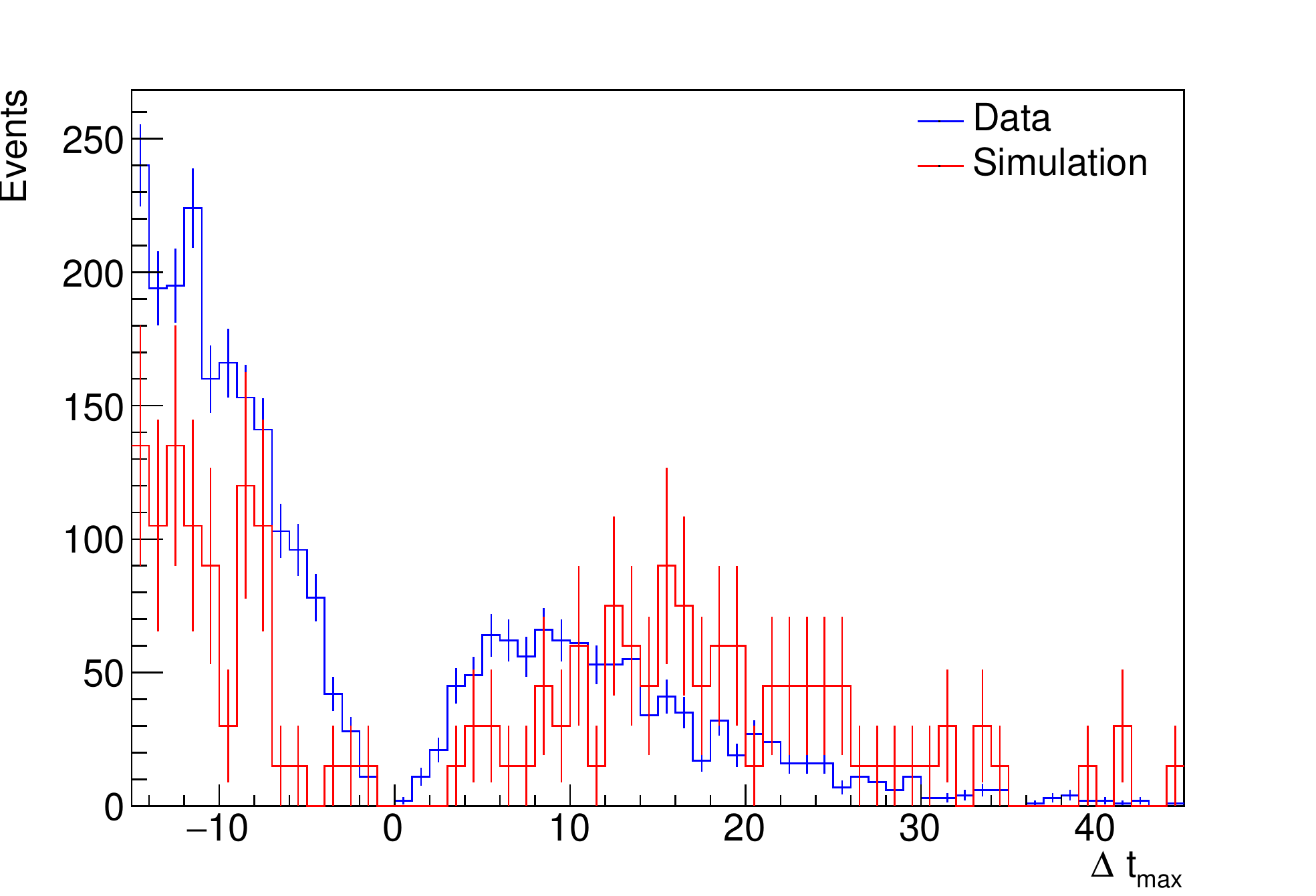}
  \caption{The $\Delta t_{\textrm{max}}$ in cosmic muon events for data (blue) and simulation (red).}
  \label{fig:cosmicSimTimeDifference}
\end{figure}

A further normalization is needed to calibrate the probability of the cosmic muon to produce a shower. 
To this end, we select events in data and Monte Carlo 
with a PMT hit in each layer, passing basic quality criteria. 
We find that the simulation needs to be scaled up by a factor of 
three in order to reproduce the rate of these events in data. After this re-scaling, we find good agreement in the 
number of scintillator bars with a detected pulse in data and simulation
(Fig.~\ref{fig:cosmicSimNbar}),
indicating that the spatial distribution and multiplicity of showers is well modelled. 
In addition, in Figures~\ref{fig:cosmicSimNPE},~\ref{fig:cosmicSimMaxMinNPE} and~\ref{fig:cosmicSimTimeDifference} we compare the modelling of the number of photoelectrons (\Npe), the ratio of the maximum to the minimum \Npe, and the $\Delta t_{\textrm{max}}$, which is defined as the maximum $|\Delta t|$ between layers with a sign then determined as positive (negative) if the layer further from (closer to) the IP has the later pulse. The tails in the $\Delta t_{\textrm{max}}$ occur from a range of sources, including random coincidence of PMT dark pulses, particles that are produced from electrons, photons and neutrons far from the detector or reflecting from the walls of the cavern, and PMT afterpulses. As will be discussed in Section~\ref{sec:signalsel}, the \Npe ratio and the $\Delta t_{\textrm{max}}$ are quantities used to define signal regions for the millicharged search. Any disagreement in the modelling of the variables shown in Figures~\ref{fig:cosmicSimNPE},~\ref{fig:cosmicSimMaxMinNPE} and~\ref{fig:cosmicSimTimeDifference} is used to define a systematic uncertainty in the relevant selection efficiency. 

A signal-like selection in the bar-detector requires only a single hit in each layer. As a result, a systematic uncertainty on the correction to the shower rate is determined by comparing the probability to pass this selection in data and simulation. The ratio of these probabilities is found to be $0.90\pm0.17$. The uncertainty is derived by taking the quadratic sum of the difference from unity with its statistical uncertainty. The scaling of the shower background is therefore taken as $3.0\pm0.6$. With the scaling applied to simulation, after requiring all signal selections detailed in Sec.~\ref{sec:signalsel}, the four-layer demonstrator yield in data is found to agree within uncertainty with the prediction from simulation.

\section{Background rejection and estimation}
\label{sec:signalsel}
The basic requirements to select signal-like events and reject backgrounds will be based on those used in 
the analysis of the demonstrator data of 
Ref.~\cite{ball2020search}.
After applying these selections the dominant remaining backgrounds are expected to be from the random overlap of PMT dark current pulses and deposits from
cosmic ray muon showers. 

\subsection{Shower background}

In what follows, the shower background is estimated from the simulation described in Section~\ref{sec:mc}, including all correction factors and uncertainties.
The first requirements to reject the cosmic background are purely geometric: a signal-like event must have exactly one hit per layer, and each hit must be in the same relative position in each layer as to form a straight path (four in line). To further reject backgrounds from cosmic showers, a veto on any deposits in the panels surrounding the bars is applied (cosmic panel veto). To reject deposits from beam muons, a veto of pulses that have an area consistent with a muon is applied to the panels at the front and back of the bar detector (beam panel veto).

After the geometric positions of the hits in the bars are determined, two further cuts reduce background. 
Since millicharged particles are expected to
deposit on average the same amount of energy in each layer,
we first require that the ratio between the maximum and minimum \Npe recorded in an event be less than 10. 

Additionally, millicharged particles would travel from the IP and would therefore be detected with a timing signature such that the layers closest to the IP are hit first. On the other hand,  showers produced by cosmic muons typically initiate above the detector,
and the bottom layers, closest to the IP, are hit last.
As a result, we require that each hit occurs within 
a narrow time window consistent with an upward-going trajectory. We require $-15$ ns $< \Delta t_{\textrm{max}} < 15$ ns, . This timing window is significantly greater than the  expected resolution of $\sim4$ ns.

We assign 10\% and 50\% relative uncertainties
on the \Npe and timing requirements, respectively.  These uncertainties are  
derived based on comparison of data from the demonstrator with simulation.

Altogether, the final expected background for the Run 3 bar detector will be $(1.2\pm0.6)\times10^{-2}$ events, and the expected background for the HL-LHC bar detector will be $(2.0\pm1.1)\times10^{-5}$ events (Table~\ref{tab:barDetTable}).
The reduction in background for the HL-LHC bar detector is due to the improved active veto provided by the significantly larger number of bars in each layer compared to the Run 3 bar detector.
    
\noindent
\begin{table}[ht]
\renewcommand{\arraystretch}{1.2}
\centering
\topcaption{Background predictions for the Run 3 and HL-LHC bar detector. The Run 3 (HL-LHC) background prediction assumes a live time of $1.5\times 10^7$ s ($3\times 10^7$ s) to accumulate 200/fb (3000/fb) of proton-proton collision data.}
\begin{tabular}{ccc}
\hline
\hline
Selection&Run 3 &HL-LHC\\
\hline
$\geq 1$ per layer &$8.1\times10^5$&$8.2\times10^7$\\
$= 1$ per layer &$6.0\times10^3$&$1.1\times10^4$\\
Cosmic panel veto &$1.1\times10^3$&$3.1\times10^3$\\
Beam muon panel veto &780&$3.0\times10^3$\\
Four in line &0.19&$2.9\times10^{-4}$\\
Max \Npe/Min \Npe $<$ 10 &0.061&$9.1\times10^{-5}$\\
-15 ns $<\Delta t_{\max}<$ 15 ns &$0.012$&$2.0\times10^{-5}$\\
\hline
\hline
\label{tab:barDetTable}
\end{tabular}
\end{table}

The signal selections made for the slab detector follow those used for the bar detector with two exceptions. The slab detector has no active panel veto, so only the self-veto on muons is used. Second, the slab geometry has a larger effective distance between the first and last layers. For low charges, pulses are only observed if the \mCP traverses the detector with a speed somewhat below $c$. To provide acceptance to such events a second timing window is considered with greater delay between layers ($15 < \Delta t_{\textrm{max}} < 45$ ns). For the Run 3 dataset the expected background for the slab detector is $7.1\pm3.9$ events for $|\Delta t_{\textrm{max}}| < 15$ ns and $1.4\pm0.8$ events for $15 < \Delta t_{\textrm{max}} < 45$ ns (Table~\ref{tab:slabDetTable}).

The selections used to reject backgrounds will be
optimized in-situ using data collected during beam off periods. The modular design of the slab detector allows for alternative layouts to be easily implemented if required.

\noindent
\begin{table}[ht]
\renewcommand{\arraystretch}{1.2}
\centering
\topcaption{Background predictions for the Run 3 slab detector. The background prediction assumes a live time of $1.5\times 10^7$ s to accumulate 200/fb of proton-proton collision data. The selections described in the last two rows define two exclusive timing categories.}
\begin{tabular}{cc}
\hline
\hline
Selection&Run 3\\
\hline
$\geq 1$ per layer &$2.0\times10^7$\\
$= 1$ per layer &$4.8\times 10^6$\\
Muon veto &$2.6\times10^5$\\
Four in line &76\\
Max \Npe/Min \Npe $<$ 10 &23\\
\hline
-15 ns $< \Delta t_{\textrm{max}} <$ 15 ns &7.1\\
15 ns $< \Delta t_{\textrm{max}}<$ 45 ns &1.4\\
\hline
\hline
\label{tab:slabDetTable}
\end{tabular}
\end{table}

\begin{figure*}[!htp]
    \centering 
    \includegraphics[width=1.5\columnwidth]{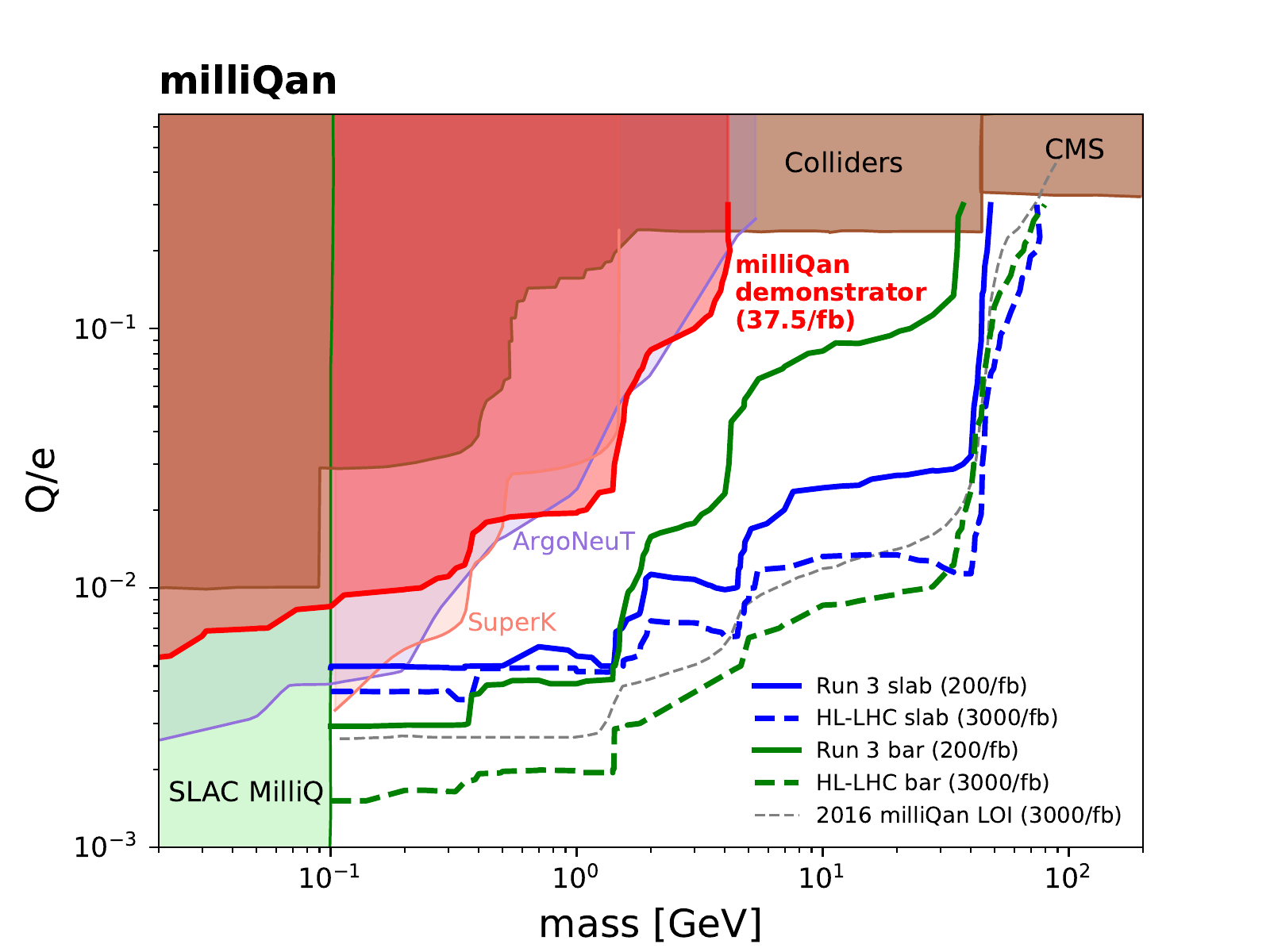}
    \caption{\protect Expected exclusion at 95\% confidence level compared to existing constraints ~\cite{Plestid_2020,MilliQ,Vogel_2014,essig2013dark,Chatrchyan_2013,Chatrchyan_2013_2,Acciarri_2020, Davidson:2000hf, Badertscher:2006fm,ball2020search}. Results from a recent reinterpretation of data collected by the BEBC WA66 beam dump experiment, which provides competitive constraints on millicharge particles with masses below a few GeV, are not shown as only 90\% confidence level contours are available~\cite{Marocco_2021}.}
    \label{fig:limit}
\end{figure*}

\subsection{PMT dark rate background}

The background from the PMT dark rate can be estimated from the typical dark rate measured during operation of the demonstrator as 2 kHz. With a trigger live time of $1.5 \times 10^7$ s, requiring a maximal time difference between layers  within 15 ns leads to a background of 0.0032 per signal-like ``path", where a path is defined as a set of 4 bars or slabs pointing
back to the IP. For the 
Run 3 bar detector, with 16 signal-like paths, this corresponds to a total background of 0.05 events, while for the slab detector, with 12 signal-like paths this background is estimated to be 0.03 events.  For the additional timing category (with $15 < \Delta t_{\textrm{max}} < 45$ ns) considered for the slab detector, the dark rate background is 0.7 events.

For the HL-LHC bar detector, assuming a live time of 
$3\times 10^7$ s, and 216 signal-like paths, the total dark rate background will be 1.4 events.  
A more precise determination of dark-rate backgrounds will be possible once the actual detector is installed.

\section{\label{sec:proj}Sensitivity projections}

The selections discussed in Section~\ref{sec:signalsel} are applied to the signal samples (simulated as discussed in Sec~\ref{sec:mc}) to determine the yield passing selection for each of the detector designs for Run 3 and the HL-LHC. Under the signal plus background hypothesis, a modified frequentist approach is used
to determine expected upper limits at 95\% confidence level
on the cross section ($\sigma$) to produce a pair of $\mCP$s, as a function of mass and charge. The approach uses the LHC-style profile likelihood ratio
as the test statistic~\cite{CMS-NOTE-2011-005} and the CLs
criterion~\cite{junk, CLsTechnique}. The expected upper limits are evaluated
through the use of asymptotic formulae~\cite{Cowan:2010js}. Figure~\ref{fig:limit} shows the exclusion at 95\% confidence level in mass and charge of the \mCP. The expected exclusion is compared to existing constraints. For Run 3, a combination of a bar and slab detector is shown to provide the strongest limits on the charge for all masses above 0.1 GeV. For the slab detector, the sensitivity is shown to be improved when the \mCP is produced near the mass threshold of a resonance. In such cases, the velocity of the \mCP is reduced and it therefore deposits more energy in the detector. At the HL-LHC, the full bar detector is shown to exceed the expected performance projected in the original milliQan letter of intent~\cite{Ball:2016zrp}, reaching charges as low as 0.0018e. A slab detector at the HL-LHC is shown to provide sensitivity for charges between 0.003 and 0.01 for all \mCP masses less than 40 GeV.

\section{\label{sec:conclusion} Conclusions}

We have reported on the expected sensitivity of detectors that we intend to install for Run 3 of the LHC. Data from the milliQan demonstrator has been used to calibrate and validate the simulation of the shower background and the simulation of the detector response. The background expected to be seen by the detectors has been estimated and the reach for millicharged particles evaluated. With a combination of a bar and slab detector, the existence of particles with mass between 10~MeV and 45~GeV could be excluded at 95\% confidence level for charges between 0.003e and 0.3e, depending on their mass. At the HL-LHC, a full bar detector is shown to extend this reach to particles with mass between 10~MeV and 80~GeV for charges between 0.0018e and 0.3e, depending on their mass. 

\section*{\label{sec:ack} Acknowledgments}

We would particularly like to thank Harvey L. Karp for his generous funding of the detector through the Harvey L. Karp Discovery Award. We congratulate our colleagues in the CERN accelerator departments for the excellent performance of the LHC and thank the technical and administrative staffs at CERN. In addition, we gratefully acknowledge the CMS Collaboration for supporting this endeavor by providing invaluable logistical and technical assistance. We thank Eder Izaguirre and Itay Yavin for their enduring contributions to this idea and Yu-Dai Tsai for valuable comments on this manuscript. Finally, we acknowledge the following funding agencies who support the investigators that carried out this research in various capacities: FWO (Belgium) under the “Excellence of Science – EOS” – be.h project n. 30820817; Swiss Funding Agencies (Switzerland); STFC (United Kingdom); DOE and NSF (USA); Lebanese University (Lebanon).

\bibliographystyle{JHEP3}
\bibliography{milliQanProjection}
\end{document}